\documentclass[%
 reprint,
superscriptaddress,
 amsmath,amssymb,
 aps,
prb,
]{revtex4-2}
\usepackage[utf8]{inputenc}
\usepackage{amsmath}
\usepackage{amsfonts}
\usepackage{amssymb}
\usepackage{graphicx}
\usepackage[caption=false]{subfig}
\usepackage{hyperref}
\usepackage{braket}
\usepackage{color,soul}
\usepackage[export]{adjustbox}
\usepackage[colorinlistoftodos]{todonotes}
\usepackage[left=2cm,right=2cm,top=2cm,bottom=2cm]{geometry}

\begin{document}

\title{Spatial noise correlations beyond nearest-neighbor in $^{28}$Si/SiGe spin qubits}
\author{J. S. Rojas-Arias}
\email{juan.rojasarias@riken.jp}
\affiliation{RIKEN, Center for Quantum Computing (RQC), Wako-shi, Saitama 351-0198, Japan}
\author{A. Noiri}
\affiliation{RIKEN, Center for Emergent Matter Science (CEMS), Wako-shi, Saitama 351-0198, Japan}
\author{P. Stano}
\affiliation{RIKEN, Center for Emergent Matter Science (CEMS), Wako-shi, Saitama 351-0198, Japan}
\affiliation{Slovak Academy of Sciences, Institute of Physics, 845 11 Bratislava, Slovakia}
\author{T. Nakajima}
\affiliation{RIKEN, Center for Emergent Matter Science (CEMS), Wako-shi, Saitama 351-0198, Japan}
\author{J. Yoneda}
\affiliation{Tokyo Institute of Technology, Tokyo Tech Academy for Super Smart Society, Tokyo 152-8552, Japan}
\author{K. Takeda}
\affiliation{RIKEN, Center for Emergent Matter Science (CEMS), Wako-shi, Saitama 351-0198, Japan}
\author{T. Kobayashi}
\affiliation{RIKEN, Center for Quantum Computing (RQC), Wako-shi, Saitama 351-0198, Japan}
\author{A. Sammak}
\affiliation{QuTech and Netherlands Organisation for Applied Scientific Research (TNO), Stieltjesweg 1, 2628 CK Delft, Netherlands}
\author{G. Scappucci}
\affiliation{QuTech and Kavli Institute of Nanoscience, Delft University of Technology, Lorentzweg 1, 2628 CJ Delft, Netherlands}

\author{D. Loss}
\affiliation{RIKEN, Center for Emergent Matter Science (CEMS), Wako-shi, Saitama 351-0198, Japan}
\affiliation{RIKEN, Center for Quantum Computing (RQC), Wako-shi, Saitama 351-0198, Japan}
\affiliation{Department of Physics, University of Basel, Klingelbergstrasse 82, CH-4056 Basel, Switzerland}
\author{S. Tarucha}
\email{tarucha@riken.jp}
\affiliation{RIKEN, Center for Emergent Matter Science (CEMS), Wako-shi, Saitama 351-0198, Japan}
\affiliation{RIKEN, Center for Quantum Computing (RQC), Wako-shi, Saitama 351-0198, Japan}

\begin{abstract}

We detect correlations in qubit-energy fluctuations of non-neighboring qubits defined in isotopically purified Si/SiGe quantum dots. At low frequencies (where the noise is strongest), the correlation coefficient reaches 10\% for a next-nearest-neighbor qubit-pair separated by 200 nm. Assigning the observed noise to be of electrical origin, a simple theoretical model quantitatively reproduces the measurements and predicts a polynomial decay of correlations with interqubit distance. Our results quantify long-range correlations of noise dephasing quantum-dot spin qubits arranged in arrays, essential for scalability and fault-tolerance of such systems.

\end{abstract}

\maketitle

\section{Introduction}

Noise is one of the biggest obstacles for the realization of quantum computing. Qubits are coupled to noisy environments that perturb the whole computational sequence: initialization, manipulation, evolution and readout. Many efforts have been made to improve each of these stages \cite{Veldhorst2014, Takeda2016, Nakajima2017, Watson2018, Zajac2018, Takeda2018, Yoneda2018, Xue2019, Nakajima2019, Yang2019, Takeda2020, Yoneda2020, Noiri2020, Blumoff2022, Xue2022, Mills2022a, Mills2022b, Noiri2022}. For example, quantum error correction (QEC) has the goal of introducing protocols to detect and correct those errors. However, its implementation is currently challenging and it requires certain thresholds for it to be feasible \cite{Rispler2020, Wang2011, VanRiggelen2022, Takeda2022}. 
{The QEC scheme to be applied depends on the characteristics of the noise and the presence of spatial correlations can be severely detrimental {\cite{Preskill2013, Aharonov2006, Clemmens2004, Szankowski2016, Paz-Silva2017, Premakumar2018, Boter2020, VonLupke2020}}.}

Decreasing noise is a challenging task that requires a good understanding of the sources and properties of the noise in the system of interest. In spin qubits defined by trapping of individual electrons in semiconductor quantum dots (QDs) \cite{Loss1998, Hanson2007}, the coherence time is limited by magnetic and/or electric noise sources. The most relevant magnetic source is the ensemble of nuclei with a finite spin in the host material, that couples to the qubit via hyperfine interaction creating a fluctuating Overhauser field \cite{Reilly2008, Malinowski2017, Tenberg2015, Chekhovich2013}. The shift to materials with a low concentration of isotopes with finite nuclear spin like Si has led to a significant improvement on coherence times \cite{Zwanenburg2013, Connors2019, Freeman2016}. Isotope engineering further suppresses decoherence due to naturally contained nuclear spins \cite{Struck2020, Yoneda2018, Wild2012, Noiri2022}, while the residual nuclear spins can still be a dominant decoherence source \cite{Dehollain2016, Collard2019, Zhao2019, Yang2019, Hensen2020}. On the other hand, electrical noise or \textit{charge noise}, couples to the qubit via the magnetic field gradient used to enable electric control of the spin \cite{Pioro-Ladriere2008, Yoneda2015, Kha2015} or through fluctuations in $g$-factor. Charge noise is one of the most intensively studied sources of noise \cite{Hooge1976}. Still, similarly to many others, a proper description is lacking.

These different sources of noise have distinct spatio-temporal properties. We would like to stress that although it is common practice to use temporal correlations to understand noise, auto-power spectra alone cannot pinpoint the noise origin. While $1/f$ spectra are usually assigned to charge noise and $1/f^2$ to nuclear, this is far from certain. Reference \cite{Struck2020} is an example where $1/f^2$ is caused by charge noise, while Ref. \cite{Jock2022} measures it as $1/f^{0.65}$. Reference \cite{Connors2022} shows a varying exponent over a wide frequency range. {The lack of consistency between studies makes it difficult to develop a proper theory. Here, we address this issue by recognizing a non-uniform spatial distribution of charge noise sources.}

{Recent experiments have observed strong spatial correlation in the noise acting on spin qubits in neighboring quantum dots \cite{Yoneda2022}. In designing the quantum computing architecture as well as better understanding the properties of the noise, it is important to investigate the length scale of this spatial correlation \cite{Preskill2013, Aharonov2006}.}

In this work we report on the measurement of spatio-temporal correlations of dephasing noise affecting non-neighboring spin qubits defined in $^{28}$Si/SiGe QDs. We find the presence of spatial correlations despite the large separation ($\approx200$ nm) between our qubits. We conjecture charge noise could be the dominant dephasing mechanism, based on the presence of cross-correlations and the expected strength of hyperfine coupling with the nuclei. Then we propose a reasonable microscopic model for charge noise deriving from an ensemble of two-level charge fluctuators. The model quantitatively agrees with the measurement results and allows us to form a clearer picture for the variability of noise spectra in spin qubit systems and predict the average scaling of correlations with interqubit distance.

The rest of this paper is organized as follows: In Sec. \ref{sec:device} we describe the device used. In Sec. \ref{sec:meas} we outline the measurement protocol and show the results on the noise spectral composition and its spatial correlation properties. Section \ref{sec:model} is focused on the postulation of the main dephasing mechanism and the theoretical model used to explain the results obtained in the previous section. Finally, we summarize our work in Sec. \ref{sec:conc}.


\section{Silicon triple quantum dot device}\label{sec:device}

\begin{figure}
	\subfloat{\includegraphics[width=0.5\columnwidth,valign=t]{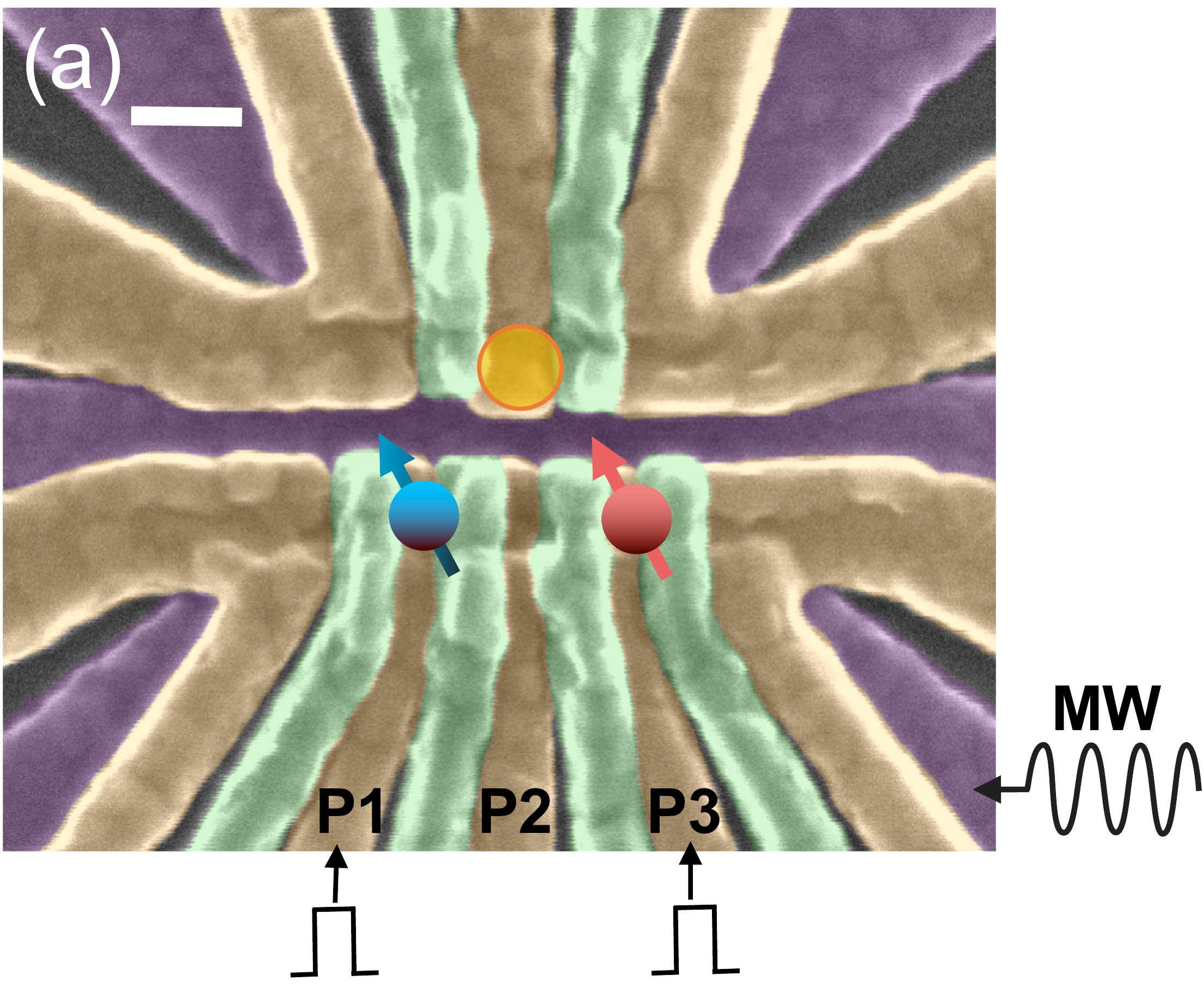}\label{fig:device}
		}
	\subfloat{\includegraphics[width=0.5\columnwidth,valign=t]{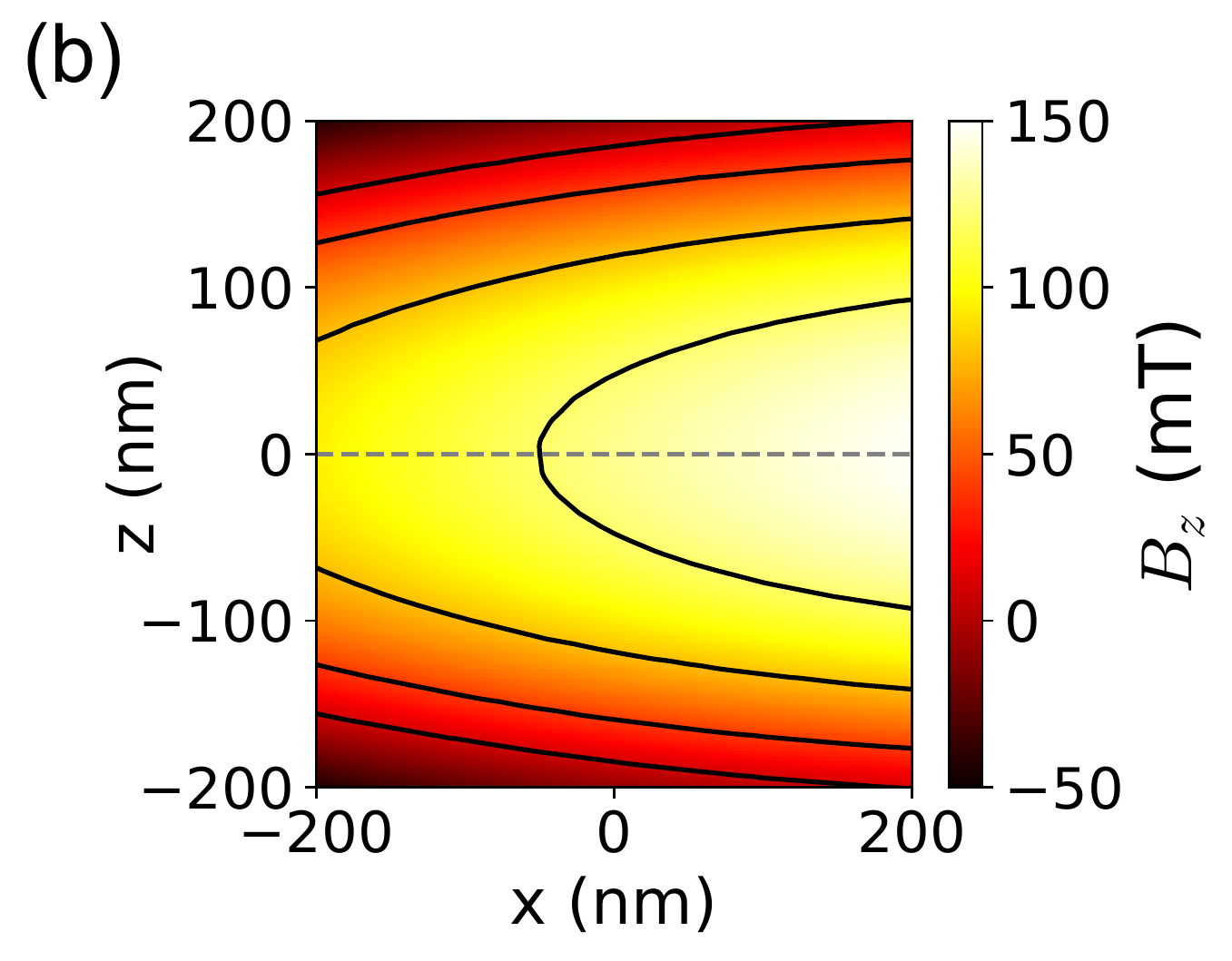}\
\label{fig:field_simulation}
		}\\
	\subfloat{\includegraphics[width=0.8\columnwidth]{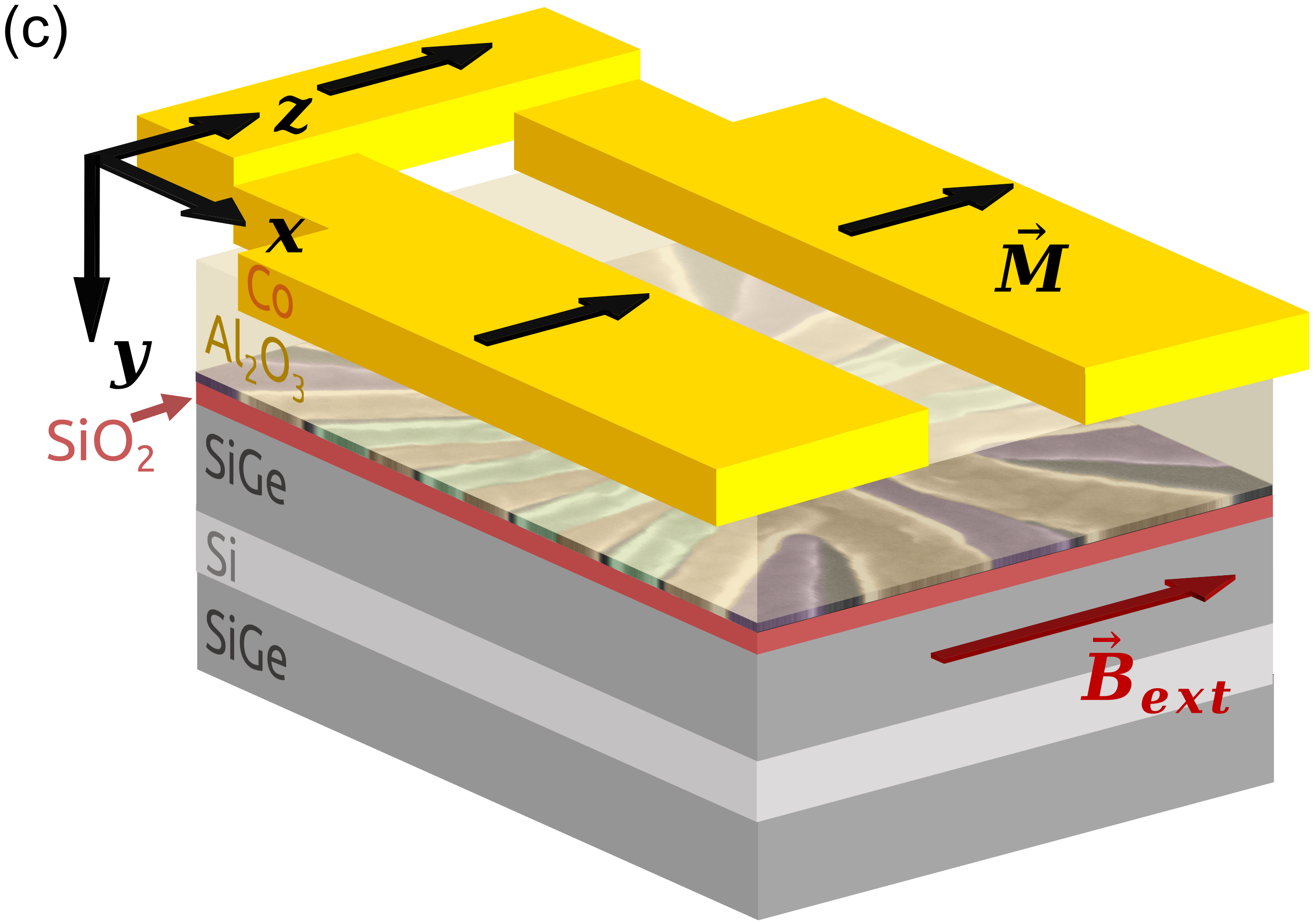}\label{fig:device2}		}
\caption{(a) False color scanning electron microscope image of a device nominally identical to the one measured. The white scale bar indicates 100 nm. (b) Simulation of the $z$ component of the micromagnet field in the QD plane. (c) Schematic depiction of the structure layer with the micromagnet (yellow) on top and a thin SiO$_2$ layer (red) right below the metallic gates. }
\label{fig:system}
\end{figure}

The triple quantum dot (TQD) device shown in Fig.~\ref{fig:device} was fabricated by patterning of overlapping aluminium gates \cite{Zajac2015} on top of an isotopically purified $^{28}$Si/SiGe wafer (remnant 800 ppm $^{29}$Si isotopes) composed of a 10 nm thick Si quantum well, a 50 nm SiGe spacer, and a 2 nm SiO$_2$ cap. We tune the device to the (1,0,1) charge configuration, meaning that a single electron is trapped in the left- and right-most dots (blue and red arrows in Fig.~\ref{fig:device}, respectively), underneath the plunger gates P1 and P3 which are separated by 180nm. The middle dot remains empty throughout the experiment. Qubits are defined in the spin states of the electrons, which are Zeeman-split in the presence of an applied in-plane external magnetic field $B_{\mathrm{ext}}=0.45$ T. With a sensor QD (orange circle in Fig.~\ref{fig:device}) in the vicinity of the TQD structure we read out the spin states by spin-to-charge conversion via energy selective tunneling \cite{Elzerman2004}, which is also used for initialization. Fast single-shot spin readout is achieved by radio-frequency reflectometry by coupling the sensor QD to a tank circuit \cite{Noiri2020}.

Electric control of the spin qubits is executed via electric-dipole spin resonance. It is enabled by the deposition of a micromagnet (MM) to create a magnetic field gradient \cite{Yoneda2015}. The MM is located on top of the device with a 30 nm thick aluminium oxide layer insulating from the metallic electrodes (Fig.~\ref{fig:device2}). A simulation of the magnetic field induced by the MM in the direction of the externally applied field is presented in Fig.~\ref{fig:field_simulation}. For the simulation, we employed the COMSOL Multipyshics\textregistered\  software with a 3D geometry following the fabrication parameters of our layered structure, including a simplified geometry for the overlapping aluminium gates. The latter (displayed in Fig.~\ref{fig:device}) consist of screening (purple), accumulation (orange) and barrier (green) gates with a thickness of 25 nm, 45 nm and 65 nm, respectively. The Rabi frequencies used for each qubit are $f^{\mathrm{Rabi}}_L=3.73$ MHz and $f^{\mathrm{Rabi}}_R=4.07$ MHz, with the subscript $L$ ($R$) referring to the left (right) qubit. The measured coherence times are {$T_{2,L}^*=6.1\ \mu$s and $T_{2,R}^*=6.9\ \mu$s} for an integration time of 100 seconds. Because the typical qubit frequency fluctuations are much larger than the measured remnant exchange coupling, $J=0.9$ kHz \cite{Noiri2022b}, the qubits are considered to be non-interacting in the following qubit control experiment.


\section{Measurement of spatio-temporal correlations}\label{sec:meas}

\begin{figure}
\centering
	\subfloat{\includegraphics[width=\columnwidth]{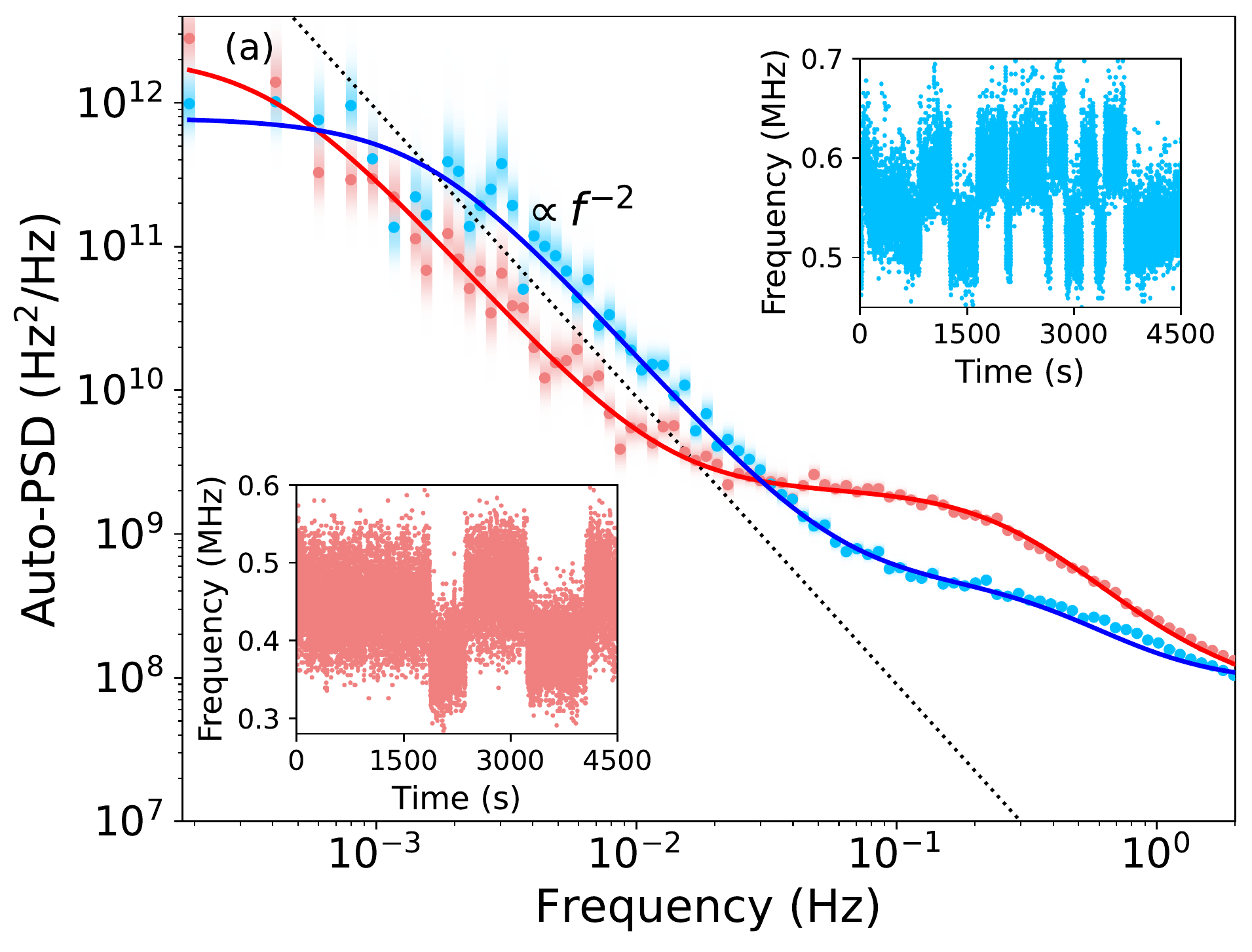}\label{fig:psd}
		}\\
	\subfloat{\includegraphics[width=\columnwidth]{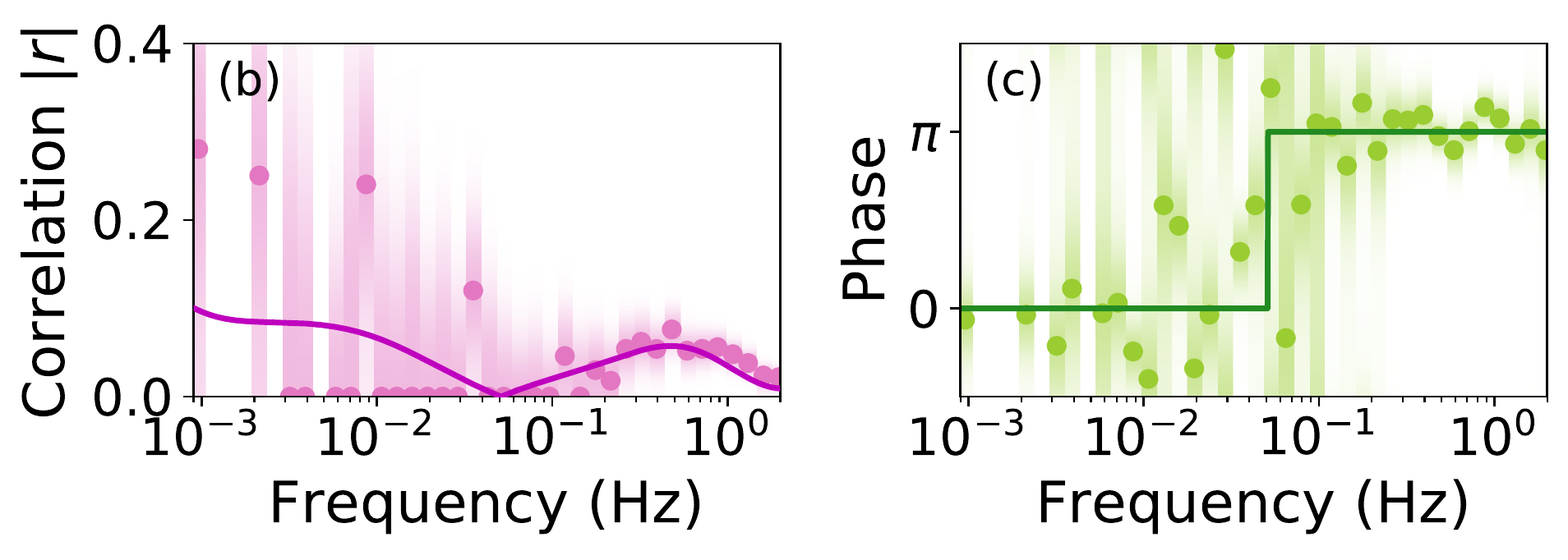}\label{fig:correlation}
		}
	\subfloat{\label{fig:phase}
		}
\caption{(a) Power spectral density of the left (blue) and right (red) qubit frequency fluctuations. The dotted line corresponds to a $f^{-2}$ dependence as a reference. Inset: sample time traces of each qubit frequency showing a 2-level switching behavior; the color code follows that of the main panel. (b) and (c) correlation coefficient amplitude and phase, respectively. In all panels, the color gradient represents the estimating distribution of the corresponding quantity at each frequency with the maximum likelihood estimator marked as points (details are discussed in Ref.~\cite{Gutierrez-Rubio2022}); the continuous lines are calculated from the distribution in Fig.~\ref{fig:charges}.}
\label{fig:meas}
\end{figure}

{In our work we study dephasing noise, which causes fluctuations in the resonant frequency of the qubits.} In order to probe the spatio-temporal correlations we measure interleaved Ramsey oscillations of the qubits repeatedly and estimate the time evolutions of their resonant frequencies using a Bayesian algorithm \cite{Delbecq2016}. An interleaved Ramsey cycle starts with the initialization of the two qubits to the spin down state. Next, a $\pi/2$ pulse, a free evolution time $t_e$ and a second $\pi/2$ pulse are applied to qubit $L$, followed by the same sequence applied to qubit $R$. The cycle is finalized by readout of the spin state of both qubits. Each cycle takes $2.09$ ms to be acquired and is repeated for $t_e$ increasing from $0.04\ \mu$s to $4\ \mu$s, in $0.04\ \mu$s steps. This repetition of cycles forms a record. A frequency value for each qubit can be extracted from a single record by Bayesian estimation \cite{Delbecq2016,Nakajima2020}. We performed $10^5$ repetitions of such records, from which we obtain the time evolution of the qubit frequencies covering a time span of $5$ hours 48 minutes. Typical samples of the time evolutions can be seen as insets in Fig.~\ref{fig:psd}. {Fast fluctuations can be seen to be modulated by a slow switching between two discrete levels in both qubits, each of them with different rates. The switching in one qubit appears to be independent from the other at first sight, but we will gain more information in our following spectral analysis.}

To analyze the spectral content of the noise, we calculate the auto power spectral densities (auto-PSD), $S_L$ and $S_R$, of the qubit frequency fluctuations $\delta\nu$ using:
\begin{equation}
C_{\alpha\beta}(f)=\int_{-\infty}^\infty
 d\tau e^{2\pi if\tau}\braket{\delta\nu_\alpha(0)\delta\nu_\beta(\tau)},
 \label{eq:PSD}
\end{equation}
where $\tau$ denotes time, $\braket{\cdots}$ a temporal average, $\alpha,\beta=L, R$ and $S_\alpha\equiv C_{\alpha\alpha}$. The cross power spectral density (cross-PSD) $C_{LR}=C_{RL}^*$ takes, in general, complex values and gives information on the spatial correlations of the noise.
It is common to relate the auto-PSDs and the cross-PSD introducing the \textit{normalized cross-PSD}, also known as \textit{correlation coefficient}:
\begin{equation}
r(f)=\dfrac{C_{LR}(f)}{\sqrt{S_L(f)S_R(f)}}.
\label{eq:correlation}
\end{equation}
The normalized cross-PSD is a complex function $r(f)=|r(f)|e^{i\phi(f)}$ whose phase provides information on the type of correlations present while its magnitude measures the proportion of correlated noise with respect to the total noise. At a given frequency, a correlation $|r|=1$ corresponds to perfect spatial correlations while $|r|=0$ means independent (uncorrelated) noise. For the phase, $\phi=0$ corresponds to positive correlations, $\phi=\pi$ to negative correlations, and other values denote a time lag between the two signals.

We calculated the auto-PSD for each qubit using Bayesian estimation of correlation functions \cite{Gutierrez-Rubio2022}, from which we are able to extract not only the values of the PSDs but also the confidence level of the estimations, the results are plotted in Fig.~\ref{fig:psd}. There we can see a Lorentzian ($\propto f^{-2}$) dependence at frequencies lower than $0.01$ Hz in both qubits. This behavior is typical of a two-level fluctuator \cite{Machlup1954}, and corresponds to the frequency switching seen in the time-traces. The normalized cross-PSD (Figures \ref{fig:correlation} and \ref{fig:phase}) in this range indicates positive correlations with a magnitude close to $0.3$, however, the level of confidence (represented as a color gradient) is too low to consider the value reliable. This highlights the importance of providing a level of confidence when reporting noise spectra and is the reason why we use the Bayesian estimation of correlation functions from Ref.~\cite{Gutierrez-Rubio2022}. Since the frequency switchings are independent from one another, noise at the lowest-frequencies ($<10^{-3}$ Hz) appears to be close to uncorrelated, but we cannot quantify the degree of correlation with reasonable confidence unless we cover a much longer time span with our measurement. The most likely origins of this noise are: hyperfine coupling to a single nuclear spin \cite{Hensen2020}, or a charge fluctuator with a long ($>$1000 s) switching time \cite{Hooge1976, McWhorter1955}.

The higher-frequency part of the spectra exhibits different slopes for each qubit and its relation with the time-traces is not straightforward. When we look at the normalized cross-PSD in this regime, we can see two interesting points. First, despite the large separation between qubits we find finite spatial noise correlations that can reach an amplitude $|r|\sim0.1$ for frequencies around 0.4 Hz. {We obtain a larger amplitude around 0.01 Hz and 0.04 Hz, albeit with a much lower confidence level.} Second, these correlations have a well-defined phase relation, with a sharp transition from positive to negative at around 0.06 Hz. The estimating distributions in the plots provide us with confidence on the measured results \cite{Gutierrez-Rubio2022}. All this confirms the correlations are beyond coincidence, proving that the resonant frequencies of these two qubits are correlated. 

So far we have seen that the measured spectral content of the dephasing noise in our qubits exhibits a non-trivial behavior, one that does not follow a typical monotonic $\propto 1/f^{b}$ dependence. Furthermore, we found the existence of cross-correlations; as high as $\sim10\%$ of the noise is correlated, even though the qubits are non-neighboring. {This can already be detrimental for QEC and raises an important question in achieving fault tolerance and scaling up the system towards a quantum computer} \cite{Clemmens2004}. In the next section, we propose a simple microscopic model for the origin of the noise, that allows us to understand the special noise spectral features measured including correlations and from which we explore the scaling of these correlations with interqubit distance.

\section{Noise as an ensemble of charge fluctuators}\label{sec:model}

We expect noise in our device to be dominated by two possible sources: the remnant $^{29}$Si nuclei, or charge noise. It is difficult to distinguish the noise origin by looking at power spectra as the one from Fig.~\ref{fig:psd}. While the most typical spectra of charge noise is $1/f^b$ with $b=1$, recent studies show that different exponents can also be obtained \cite{Struck2020, Connors2022, Elsayed2022}. On the other hand, dynamics of the nuclear spins are usually modelled as a diffusive process from which $b=2$ would be expected. However, subdiffusive behaviour has also been observed \cite{Delbecq2016, Nakajima2020}. In this section we conjecture on the dominant dephasing mechanism and introduce our theoretical model for it.

From the presence of cross-correlations we are able to gain some insight \cite{Yoneda2022}. While nuclear noise is local, limited by the extent of the electron's wavefunction (see Appendix \ref{app:nuclear}), charge noise from fluctuating electric fields should have a longer range leading to spatial correlations. The fact that we are able to resolve finite correlations in our distant qubits, confirms the presence of charge noise at least on a measurable level. Furthermore, for the typical size of our QDs ($\sim28$ nm in diameter for a 2 meV harmonic confinement) the maximum hyperfine coupling strength achievable for a single nuclear spin is $\sim10$ kHz (calculated with Eq.~\eqref{eq:shift}). Thus, hyperfine interaction is about one order of magnitude smaller than the two-level frequency switching seen in the time-traces in Fig.~\ref{fig:psd}.
Hence, this suggests that low-frequency noise in our device is not dominated by nuclear spin noise, just as one would expect for a device made from isotopically purified silicon with a micromagnet \cite{Yoneda2018}. Although we cannot fully rule out the effect of nuclear spins, especially at intermediate frequencies (see Fig.~\ref{fig:nuclear}), on what follows we assume charge noise is the dominant dephasing mechanism and see whether it can account for the measured noise spectra.

Several studies \cite{Kuhlmann2013, Connors2019, Elsayed2022} point at fluctuating electric fields possibly having their origin in charge two-level systems (TLSs) located in the oxide layer as the one shown in red in Fig.~\ref{fig:device2}. 
The noise spectrum of a TLS is a Lorentzian. It is believed that $1/f$-like noise spectra arise from a superposition of many ($\rightarrow\infty$) Lorentzians with a certain distribution of switching rates \cite{Hooge1976}. This is known as \textit{McWhorter's model} \cite{McWhorter1955}.

In our model, instead of assuming a large amount of TLSs with a specific distribution of switching times we notice that just a few fluctuators in the oxide are sufficient to reproduce our experimental results. This is, at the same time, more physical since it is hard to reconcile the idea of an infinite number of fluctuators with our nanometre sized dots and the non-uniform spectral features of Fig.~\ref{fig:psd}. Spatial correlations naturally arise from our model due to the common origin of the noise affecting all dots.

Fluctuating electric fields $\delta \vec{E}$ affect the qubits by shifting the position of the QDs within the MM field shown in Fig.~\ref{fig:field_simulation} creating, in turn, a fluctuating qubit frequency leading to dephasing. These fluctuations in the qubit frequencies can be written to first order in the electric field as:
\begin{equation}
\delta\nu_\alpha=\dfrac{g\mu_B}{h}\dfrac{e}{m\omega_\alpha^2}\left(\dfrac{\partial B_\alpha}{\partial x}\delta E_\alpha^x+\dfrac{\partial B_\alpha}{\partial z}\delta E_\alpha^z\right),
\label{eq:freq_fluc}
\end{equation}
where $\hbar\omega_\alpha$ is the confinement energy, $B_\alpha$ the $z$ component of the local magnetic field at QD $\alpha$, $\mu_B$ Bohr's magneton, $e$ the electron charge, $m$ its effective mass, $g$ the g-factor and $h$ Planck's constant. Position shifts in the $y$ direction, perpendicular to the QD plane, are neglected due to the strong confinement. 

We write the fluctuating electric fields from Eq.~\eqref{eq:freq_fluc} as a superposition of charge fluctuators $\delta q_i$ coupled to the qubit electron via Coulomb interaction:
\begin{equation}
\delta\vec{E}_\alpha=\sum_i\frac{1}{4\pi\epsilon\epsilon_0}\frac{\vec{x}_\alpha-\vec{x}_i}{|\vec{x}_\alpha-\vec{x}_i|^3}\delta q_i,
\label{eq:field_fluc}
\end{equation}
with $\epsilon$ the effective dielectric constant of the layered structure, $\epsilon_0$ the permittivity of vacuum, $\vec{x}_\alpha$ the position of the QD $\alpha$ and $\vec{x}_i$ that for fluctuator $i$. We assume independent fluctuators, that is, $\braket{\delta q_i(t)\delta q_j(t+\tau)}=(e^2/4)\delta_{ij}\exp(-\tau/t_i)$, where $t_i$ is a characteristic switching time for each TLS. Given a distribution of fluctuators and the location of the QDs, Eqs.~\eqref{eq:PSD}, \eqref{eq:freq_fluc} and \eqref{eq:field_fluc} allow us to calculate the auto-PSDs and the cross-PSD, from which we can obtain the correlation factor by its definition in Eq.~\eqref{eq:correlation}.

\begin{figure}
\centering
\includegraphics[width=\columnwidth]{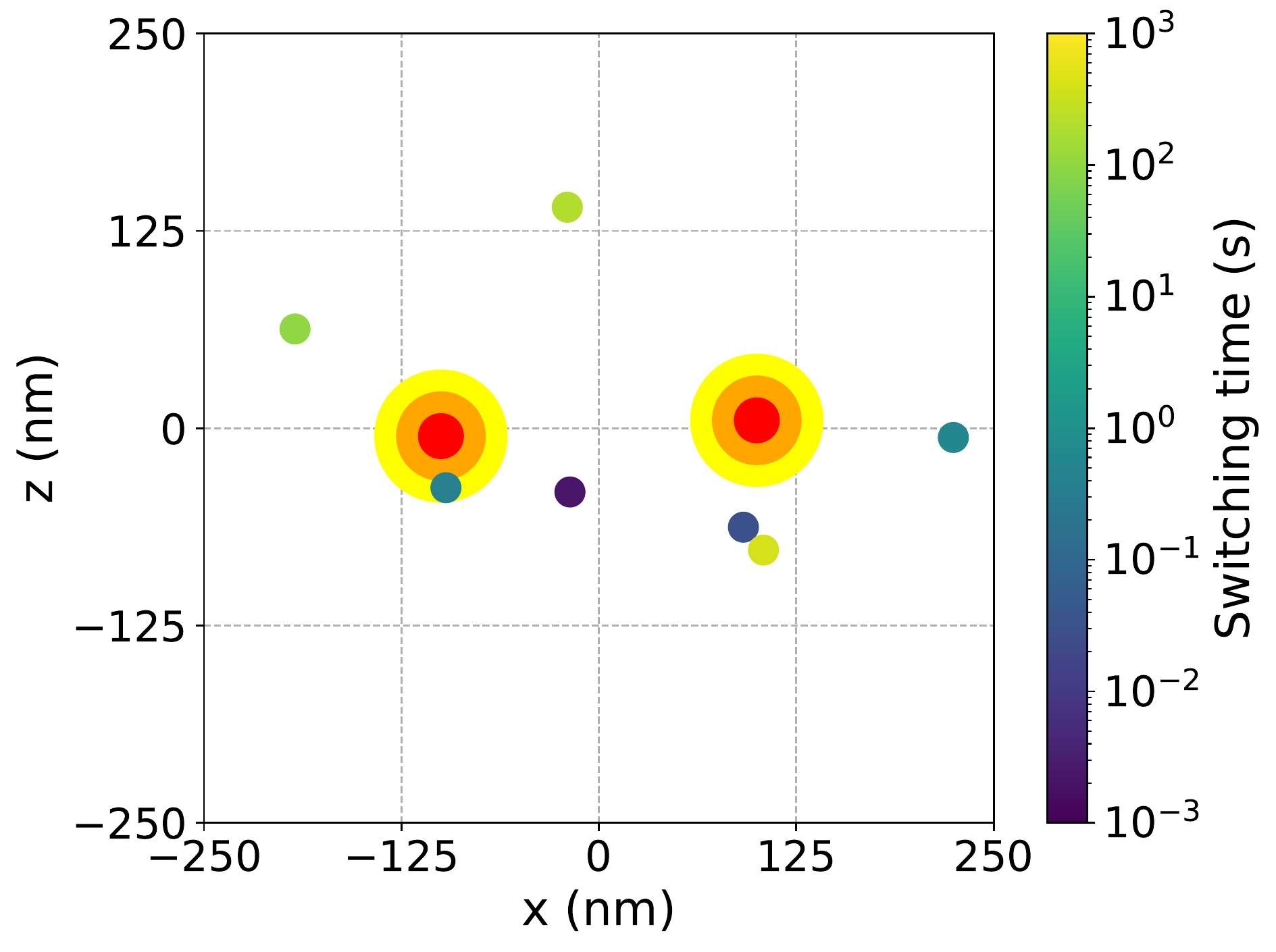}
\caption{One possible set of charge fluctuators that reproduces the experimental data. The tonality gives the switching time of the corresponding fluctuator. All charges are located 50 nm above the QD plane and have an opposite image charge 4 nm higher. The two disks in the center correspond to the QDs with yellow, orange and red representing 3, 2 and 1 standard deviations of the electron's wavefunction, respectively, for a 2meV parabolic confinement. The QDs are located at $(x,z)=(\pm100\ \mathrm{nm},\pm5\ \mathrm{nm})$. We used $m=0.2m_0$ with $m_0$ the free electron mass and $\epsilon=13$.}
\label{fig:charges}
\end{figure}

We use a least squares minimization algorithm to find positions of TLSs in space and their switching times, fitting to the correlation spectra. We increase the number of TLSs until a satisfactory fit is realized. For simplicity, we keep all the TLSs in the same plane located at the interface between the SiGe spacer and the SiO$_2$ layer, with coordinate $y=-50$ nm (the QD plane being at $y=0$). This does not have to be the case and models linking the switching time with the in-oxide depth are possible \cite{Weissman1988}. However, {they do not provide relevant information to our study}. To account for screening due to the metallic gates \cite{Zajac2016}, for each fluctuator we include an image
charge at $y=-54$ nm, as a ``reflection'' through a metal plane 2 nm above the fluctuators. Again, for the sake of simplicity, we limit ourselves to one image per fluctuator, but the treatment can be extended to include more images to account for higher order effects at the interfaces between dielectrics.
The values of the magnetic field gradients are obtained from Fig.~\ref{fig:field_simulation} and the QD positions. Although we do not know precisely the location of the QDs, the measured anti-correlations suggest that the qubits could be subject to opposite magnetic gradients. {Therefore, as an example, in our calculation we locate them at opposite sides of the $z=0$ line in }Fig.~\ref{fig:field_simulation}.

Figure \ref{fig:charges} depicts a possible distribution of TLSs in the oxide that leads to the curves in Fig.~\ref{fig:meas}. The spectral content of the noise and the type and strength of correlations can be quantitatively accounted for by the model with only seven fluctuators, corresponding to a density $n_c\sim10^{10}\ \mathrm{cm}^{-2}$. We point out the significance of screening when obtaining this value; if we remove the image charge from our model, the effect of even a single TLS near the vicinity of the dots is already about two orders of magnitude larger than the measured auto-PSDs.

\begin{figure}
\centering
\includegraphics[width=0.9\columnwidth]{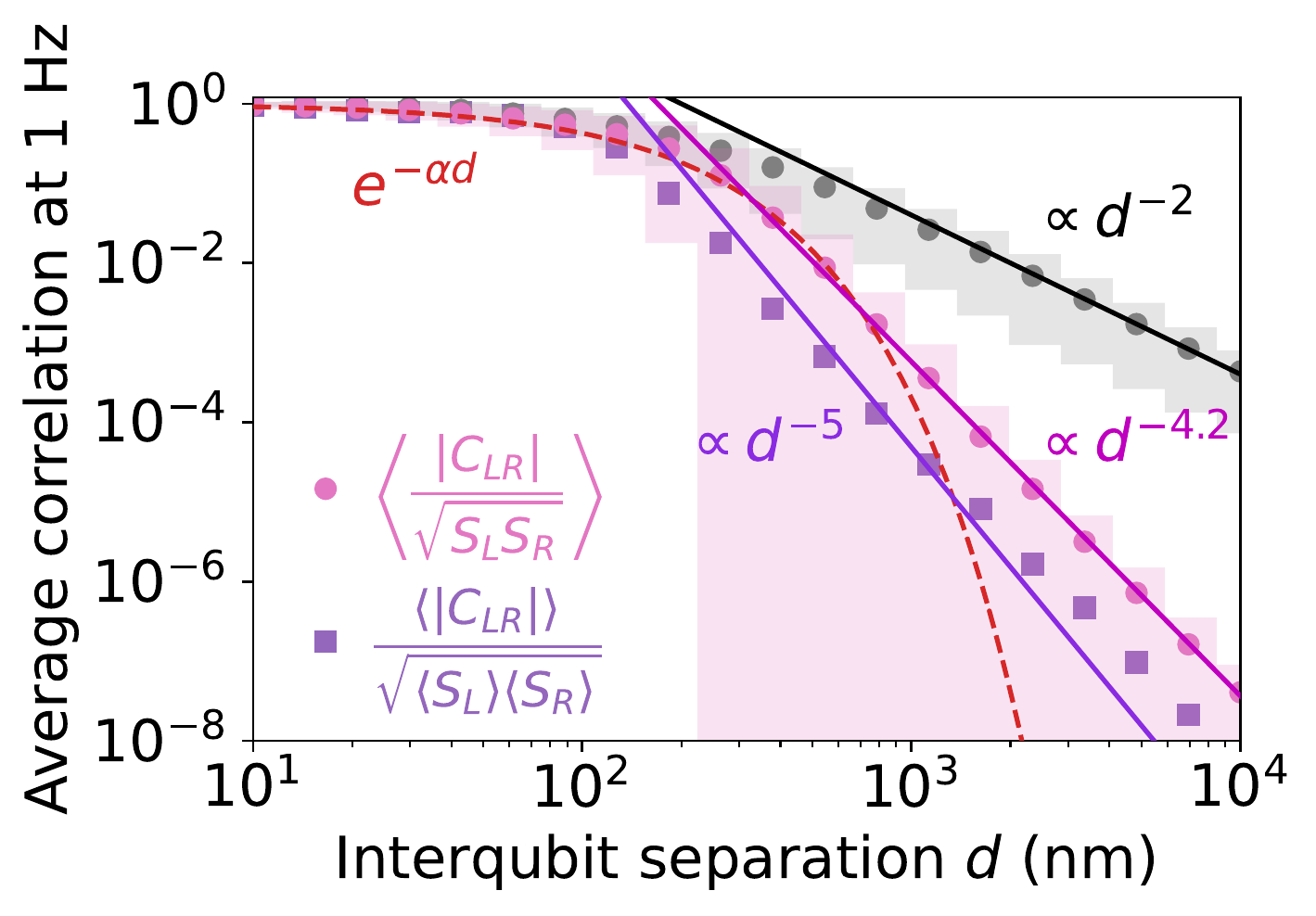}
\caption{Average normalized correlation amplitude at 1 Hz as a function of the interqubit separation. Pink (gray) circles correspond to the simulation of $\braket{r}$ with (without) screening due to the metallic gates. Shaded regions represent one standard deviation.  Purple squares correspond to $\bar{r}$. For reference, the red dotted line corresponds to an exponential decay $\exp(-\alpha d)$ with $\alpha=0.0085$ nm$^{-1}$, while the magenta (black) curve is proportional to $d^{-4.2}$ ($d^{-2}$). The purple line is $\propto d^{-5}$ following the result from Appendix~\ref{app:integral}.}
\label{fig:scaling}
\end{figure}

Perhaps more crucial than the magnitude of the noise correlations is their scaling with interqubit separation \cite{Preskill2013, Aharonov2006}. We take advantage of our model to investigate this scaling by simulation of qubits with increasing separation. For these simulations we locate the qubits at $(x,z)=(\pm d/2,0)$ and assign fixed magnetic gradients $\partial B/\partial x=0.1$ mT/nm and $\partial B/\partial z=0$ for both qubits, irrespective of their separation $d$. We generate 3000 random distributions of TLSs in a simulation square-shaped area of 35 $\mu$m side length, with switching times in the range $(10^{-5},\ 10^5)$ s, using linearly and logarithmically uniform distributions for space and time, respectively. We note, however, that our ensemble-averaged results for the normalized cross-PSD are independent of the specific distribution of switching times (see Eq.~\eqref{eq:rbar} and the comment beneath it). The density of fluctuators is set to $n_c=10^{10}$ cm$^{-2}$ as previously determined and remaining parameters are kept the same. Each distribution yields different spectral characteristics (see Appendix \ref{app:scaling}), making the PSDs not self-averaging \cite{Schriefl2006}, so we calculate the average and standard deviation from the whole ensemble of generated distributions. We evaluate two quantities: the average normalized cross-PSD magnitude $\braket{r}\equiv \braket{|C_{LR}|/\sqrt{S_L S_R}}$, and the average cross-PSD magnitude normalized by the average auto-PSDs $\bar{r}\equiv\braket{|C_{LR}|}/\sqrt{\braket{S_L}\braket{S_R}}$, introduced for comparison with analytic results. We record the values at 1 Hz for different qubit separations and display the results in Fig.~\ref{fig:scaling}. The average correlation $\braket{r}$ initially appears to decay exponentially, but eventually a polynomial tail $\propto d^{-4.2}$ is observed at large interqubit separations. For $\bar{r}$ we observe smaller values that follow a similar polynomial tail. To highlight the effect of screening, we perform the same simulation without the image charges and obtain the gray circles, which show a much slower decay $\braket{r}\propto d^{-2}$. We note that these averages calculated from random sampling show a trend but may require larger ensembles to ensure good convergence. An analytic result, however, can be obtained for $\bar{r}$. The derivation is presented in Appendix~\ref{app:integral} and yields $\bar{r}\propto d^{-5}+\mathcal{O}(d^{-7})$ shown as a purple curve in Fig.~\ref{fig:scaling}. The slight disagreement with the simulation suggests that the random sampling has not reached convergence. Given the similar polynomial decays, it is likely that $\braket{r}$ also shows a decay $\propto d^{-5}$ but we could not obtain analytic results for it. Nevertheless, given the power-law decaying spatial correlations, screening becomes essential for such correlations to decay fast enough to enable a reliable quantum computation \cite{Aharonov2006}.

We mention a few things that can be learned from the model. First,
charge noise properties reflect a specific distribution of fluctuators which is device---and potentially also cooldown---dependent. Unless these sources reach some kind of self-averaging statistical limit, a single set of PSDs from a given device will not fit any generic theory predicting universal behavior. Second, the distribution of charges leading to particular spectral characteristics is not unique (see Appendix \ref{app:distribution}). Third, while each TLS affects both qubits and individually induces perfectly correlated noise, spatial correlations will decrease upon adding noise from a collection of fluctuators with different couplings \cite{Boter2020}. This, perhaps counter-intuitive, behavior is shown in Fig.~\ref{fig:scaling_density}, where high densities of TLSs serve to decrease the degree of correlation in the noise. 
Fourth, while we considered charge traps in the oxide, defects in the quantum well \cite{PaqueletWuetz2022} could contribute as well. In this case, the TLSs should be dipoles as a charge trap distant from the screening metal gates would cause a noise magnitude inconsistent with our measurements. 
Nevertheless, our results remain unaltered because the description of a charge trap close to a screening plane is equivalent to that of a fluctuating dipole. 
Finally, screening plays a major role in both the magnitude and range of the noise. It is advised to have a high density of metallic gates close to the substrate surface to screen the effect of the fluctuators, favoring overlapping gate designs as the one used in this study. Even though we measured correlated noise beyond nearest-neighbor qubits, screening could make spatial correlations decay fast enough with distance such that QEC remains feasible.

\section{Summary}\label{sec:conc}

We measured the auto- and cross-correlation noise spectra of non-neighboring spin qubits defined in $^{28}$Si/SiGe quantum dots by simultaneous measurement of Ramsey oscillations. The obtained low-frequency ($\sim10^{-3}$ Hz) noise appears to be uncorrelated and exhibits a Lorentzian shape that arises from coupling to a two-level system. Our main result lies in the higher-frequency part of the spectra, where we find well-defined cross-correlations, with a magnitude about $10\%$, despite the qubits being non-neighboring. We were able to quantitatively reproduce the measured data by modeling the noise as a superposition of a few independent charge two-level systems located in an oxide layer of the device. Our model explains the variability in charge noise spectra and highlights the importance of screening not just to decrease the noise magnitude but also its range. We predict a polynomial decay of noise cross-correlations with interqubit separation which could be checked in future measurements in larger spin qubit arrays. Our study motivates further research on noise spatial correlations, not only in spin qubit systems but in any candidate for quantum computing. Noise correlation properties are crucial for the implementation of quantum error correction and are another important benchmark to take into account.

\appendix
\section{Spatial correlations due to nuclear spin noise}\label{app:nuclear}

Here we explain how the nuclear spin noise cannot cause the observed correlations. We do so by calculating the auto- and cross-PSD via a diffusive model. The frequency shift $\delta\nu$ experienced by a qubit with spin $\vec{S}$ due to its interaction with a set of nuclear spins $\vec{I}_n$ can be written as:

\begin{equation}
\delta\nu = \dfrac{v_0A}{h}\sum_n|\psi(\vec{x}_n)|^2I_n^z,
\label{eq:shift}
\end{equation}
where  $\psi(\vec{x})$ is the electron's wavefunction, $v_0$ the unit cell volume and $A=2.4\ \mu$eV the hyperfine coupling strength for $^{29}$Si \cite{Philippopoulos2020,Assali2011,Schliemann2003}. We replace the discrete summation over nuclei by an integral over space introducing the (dimensionless) nuclear polarization density $P(\vec{x},t)$:

\begin{equation*}
\delta\nu=\dfrac{pA}{h}\int d^3x\ P(\vec{x},t)|\psi(\vec{x})|^2,
\end{equation*}
with $p$ the fraction of $^{29}$Si nuclei.

The dynamics of the nuclei with their magnetic dipole-dipole interaction show a diffusion-like behavior in the absence of nuclear spin relaxation \cite{Tenberg2015, Malinowski2017, Reilly2008, TaylorThesis}. Thus, we write the evolution of the nuclear polarization as:

\begin{equation*}
\dfrac{\partial P(\vec{x},t)}{\partial t}=D\nabla^2P(\vec{x},t).
\end{equation*}
The stochastic nature of the nuclear spins is grasped by inserting a stochastic force $\xi(\vec{x},t)$ into the diffusion equation for $P$. Moving to Fourier space this leads to:

\begin{equation*}
\dfrac{\partial P_{\vec{k}}(t)}{\partial t}=-4\pi^2k^2DP_{\vec{k}}(t)+\xi_{\vec{k}}(t).
\end{equation*}
Solving the diffusion equation we obtain for an unpolarized system:

\begin{equation}
P_{\vec{k}}(t)=\int_0^\infty dt'\ \xi_{\vec{k}}(t-t')e^{-4\pi^2k^2Dt'}.
\label{eq:polarization}
\end{equation}

Now, we are interested in calculating the cross-correlator $C_{LR}(t,t')\equiv\braket{\delta\nu_L(t)\delta\nu_R(t')}$. To do so, we assume that the statistics of the random forces are independent in space and time, i.e.:

\begin{equation}
\braket{\xi_{{\vec{k}}_1}(t_1)\xi_{{\vec{k}}_2}(t_2)}=\delta(t_1-t_2)\delta({\vec{k}}_1+{\vec{k}}_2)\Xi(k_1).
\label{eq:random_force}
\end{equation}

\begin{figure}
\centering
\includegraphics[width=\columnwidth]{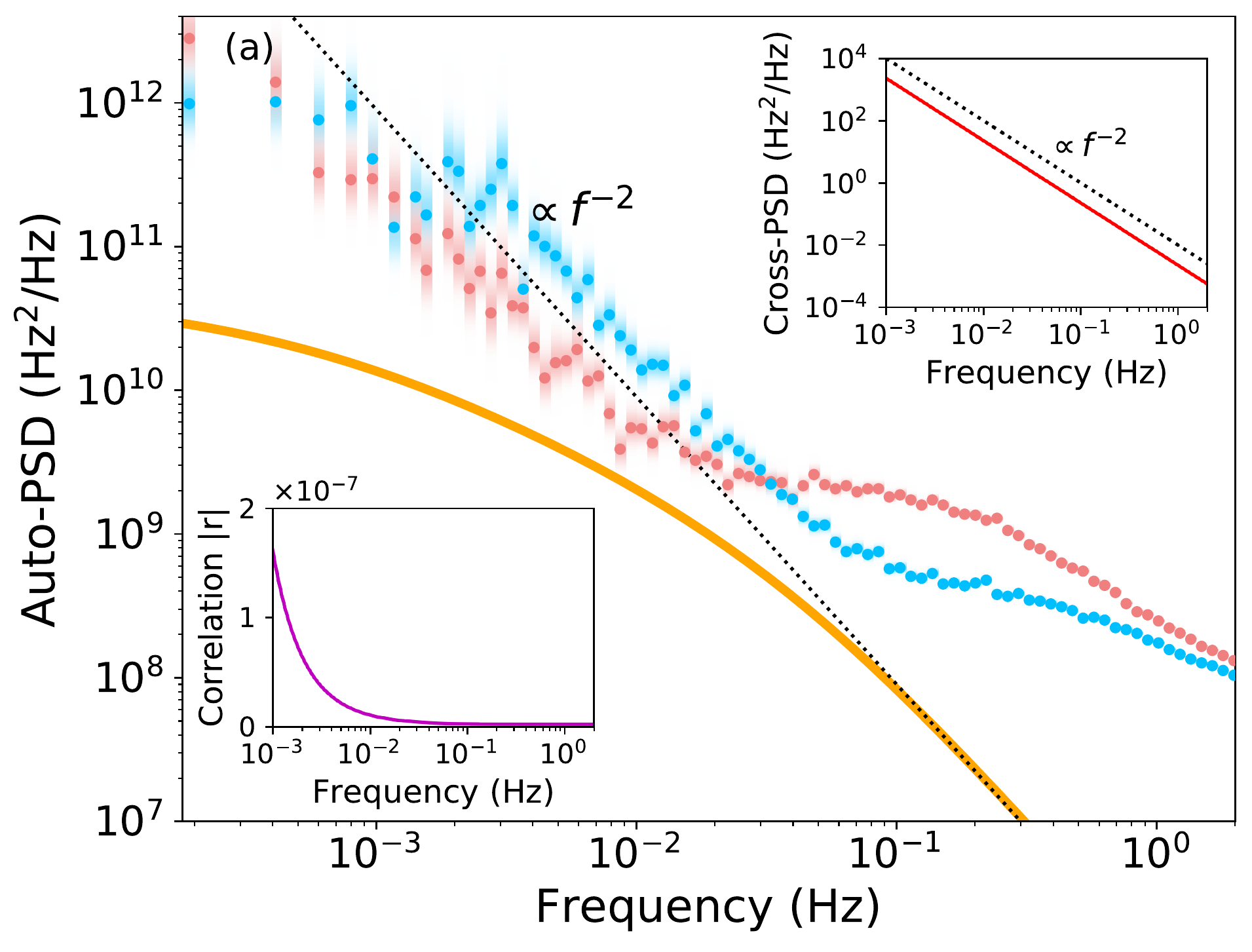}
\caption{The main panel of Fig.~\ref{fig:psd} including the auto power spectral density of the nuclear spin noise (orange) calculated according to Eq.~\ref{eq:nuclear_correlator} for $d=0$. Insets: cross power spectral density and correlation coefficient calculated for an interdot separation of 200 nm. The dotted lines are for reference.}
\label{fig:nuclear}
\end{figure}

From Eqs.~\eqref{eq:polarization} and \eqref{eq:random_force} we get the following cross-correlator:

\begin{align}
C_{LR}(t,t')=&pA^2\dfrac{I(I+1)}{3} v_0\nonumber\\
&\times\int d^3k\ |\psi_L|^2_{\vec{k}}|\psi_R|^2_{-{\vec{k}}}\ e^{-4\pi^2k^2D|t-t'|},
\label{eq:correlator_k}
\end{align}
where we evaluated $\Xi(k)=8\pi^2k^2DI(I+1) v_0p/3$ by direct calculation of $C_{LL}(t,t)$ from Eq.~\eqref{eq:shift}. The cross correlator from Eq.~\eqref{eq:correlator_k} can be evaluated if we assume a Gaussian form for the electron wavefunction:

\begin{equation*}
|\psi_\alpha(\vec{x})|^2=\dfrac{1}{\pi^{3/2}l_p^2l_y}\exp\left(-\dfrac{(x-x_\alpha)^2+z^2}{l_p^2}-\dfrac{y^2}{l_y^2}\right),
\end{equation*}
with $l_p$ and $l_y$ the spreads in-plane and out-of-plane, respectively. Using this we get the final result for the cross-correlator:

\begin{align}
C_{LR}(t,t')=&\frac{C_0\exp[-\kappa/(1+\gamma|t-t'|)]}{(1+\gamma|t-t'|)(1+\gamma\zeta|t-t'|)^{1/2}},
\label{eq:nuclear_correlator}
\end{align}
where we defined $\gamma=2D/l_p^2$, $\kappa=2d^2/l_p^2$, $\zeta= l_p^2/l_y^2$, $2d=x_R-x_L$ and:

\begin{equation}
C_0=p\dfrac{A^2}{\sqrt{2\pi}}\dfrac{I(I+1)}{3}\dfrac{a_0^3}{8V_D},
\end{equation}
with $a_0$ the lattice constant of silicon and $V_D=2\pi l_p^2l_y$ the QD volume. The Fourier transform of Eq.~\eqref{eq:nuclear_correlator} gives us the cross-PSD from which we can evaluate the spatial correlations. The auto-PSD is obtained by setting $d=0$ and it is plotted in Fig.~\ref{fig:nuclear} for $D=0.4$ nm$^2/$s \cite{Hayashi2008} and a typical QD size of $(l_p,l_y)=(14\ \mathrm{nm},3\ \mathrm{nm})$. There we can see that the nuclear noise auto-PSD is lower than the measured values, giving further indication that the main dephasing mechanism is charge noise. The correlation coefficient is $|r|<10^{-6}$ in the frequency range of interest as shown in the insets of Fig.~\ref{fig:nuclear}, a value much smaller than that displayed in Fig.~\ref{fig:correlation}. This very small value occurs because the Overhauser field experienced by the qubits is determined by the spread of their respective wavefunction; if there is not a considerable overlap, the local nuclear dynamics evolve rather independently. Thus, nuclear spin noise is highly uncorrelated, especially at the large interdot separation in our device.

\section{Scaling of cross-correlations with distance}\label{app:scaling}

\begin{figure}[htb]
\centering
\includegraphics[width=0.9\columnwidth]{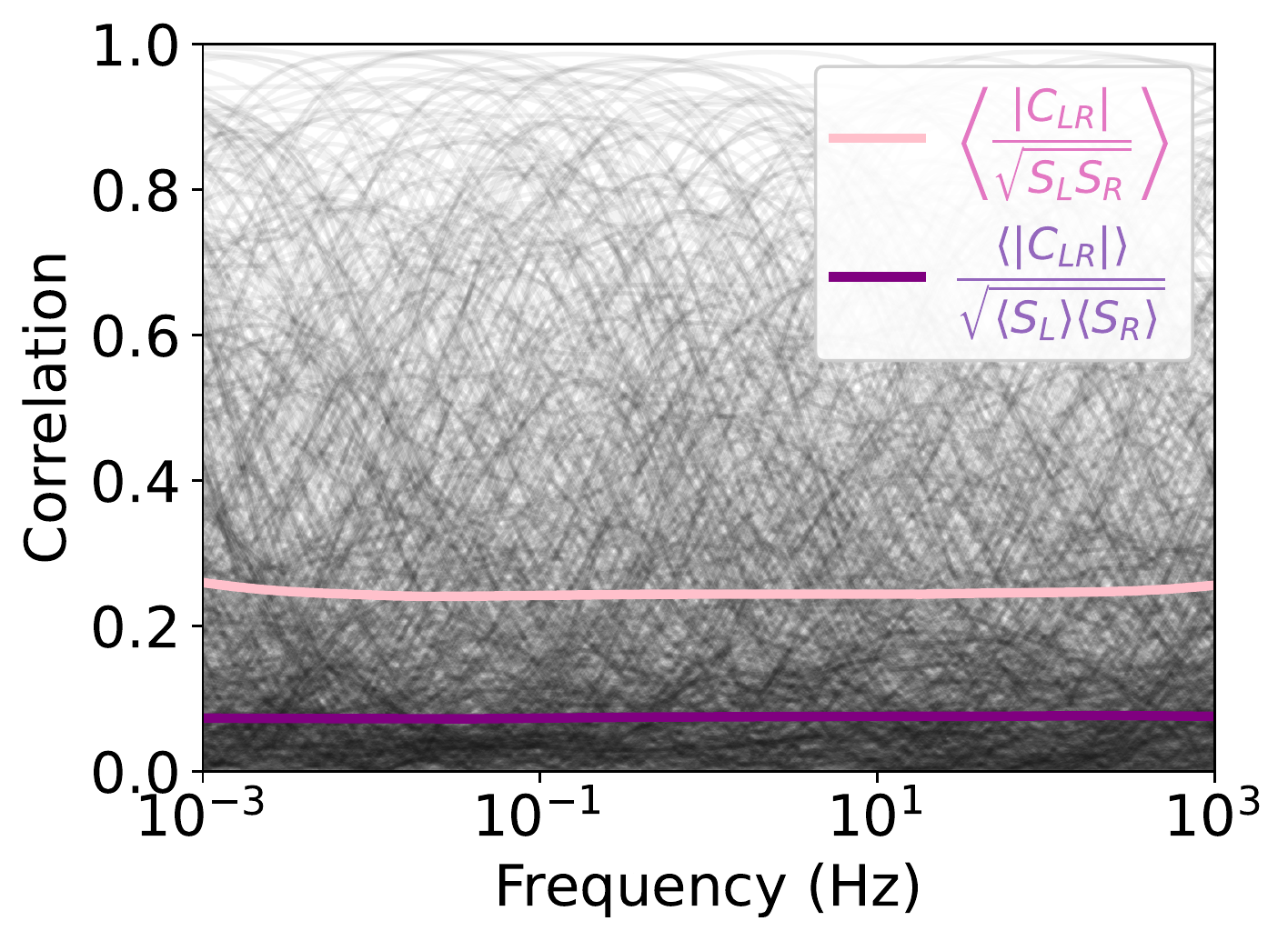}
\caption{Normalized cross-PSD amplitude calculated from different distributions of fluctuators for a separation of 200 nm between qubits. The black lines correspond to the results of 1000 different ensembles with $\braket{r}$ in pink and $\bar{r}$ in purple.}
\label{fig:corr_ensemble}
\end{figure}

When we simulate random ensembles of TLSs, the noise spectral characteristics vary for each distribution. Namely, the result is sensitive to the properties of a few TLSs, making it not self-averaging \cite{Schriefl2006}. This is seen in Fig.~\ref{fig:corr_ensemble}, where the normalized cross-PSD can take basically any value between uncorrelated and perfectly correlated noise. On the other hand, the average (be it $\braket{r}$ or $\bar{r}$) takes on an almost constant value independent of frequency (any small features seen probably arise from our finite sampling). This allows us to extract a mean correlation strength for a certain interqubit separation. For convention, we take the value at 1 Hz but the scaling with distance is independent of this choice.

\begin{figure}[h]
\centering
\includegraphics[width=0.9\columnwidth]{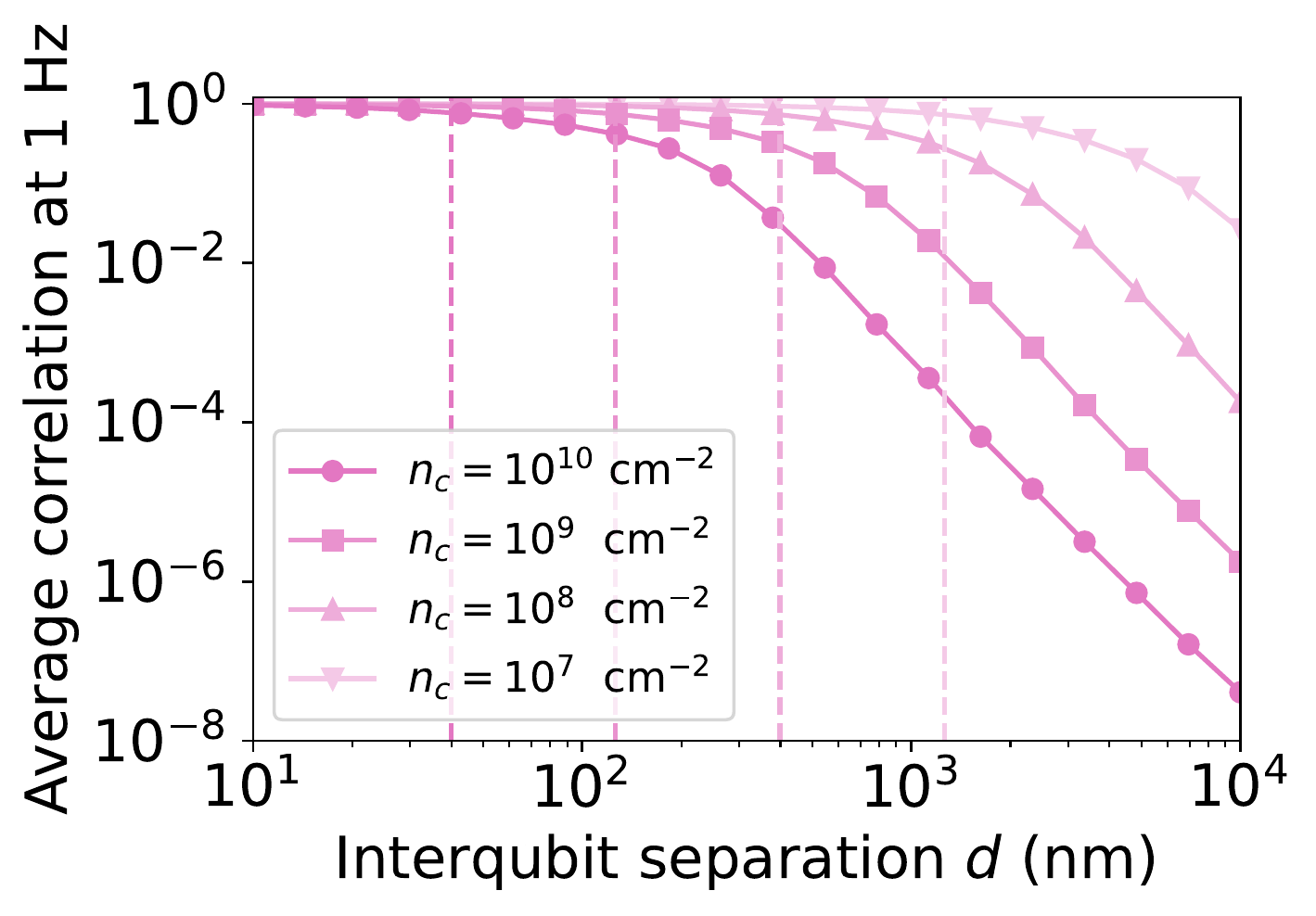}
\caption{Average normalized cross-PSD amplitude $\braket{r}$ at 1 Hz for different densities of TLSs. Vertical dashed lines signal the
mean distance between fluctuators equal to $1/\sqrt{2\pi n_c}$.}
\label{fig:scaling_density}
\end{figure}

When the qubits are further apart, strong cross-correlations become less likely and the average $\braket{r}$ decreases. The interqubit separation at which correlations start to decay is determined by the mean distance between TLSs as is shown in Fig.~\ref{fig:scaling_density}. At higher densities, the fluctuators are closer to each other such that it becomes more probable for the two qubits to strongly couple to a different group of fluctuators, leading to small noise cross-correlations. Although the average correlation magnitude decreases with increasing density of charge fluctuators, the overall noise amplitude increases proportionally.

\section{Analytical calculation of $\bar{r}$}\label{app:integral}

Here, we present an analytical calculation of $\bar{r}=\braket{|C_{LR}|}/\sqrt{\braket{S_L}\braket{S_R}}$ to determine the scaling of cross-correlations with interqubit separation.

We start by treating each charge TLS with its image as an electric dipole, with fluctuating moment $\delta\vec{p}=\delta q\vec{l}=\delta ql\hat{y}$ and a position $y$ component $y_0=-52$ nm, where the qubits are at $y=0$ and $l=4$ nm following the parameters of our device. In this way, the $x$ component of the electric field at the position $\vec{x}_\alpha$ of qubit $\alpha$ due to a single fluctuator-image pair $k$, located at $\vec{x}_k=(x_k,y_0,z_k)$, can be approximated as:
\begin{align}
\delta {E}_{x,k}^\mathrm{dip}(\vec{x})\approx\delta q_k\frac{3y_0l}{4\pi\epsilon\epsilon_0}\frac{{x}_\alpha-{x}_{k}}{[(x_\alpha-x_k)^2+y_0^2+z_k^2]^{5/2}}.
\end{align}
With this, the general correlator of electric fields can be written as:
\begin{align}
C_{E_\alpha^xE_\beta^x}(f)=\chi_0&\sum_{k=1}^{N_\mathrm{TLS}}\frac{({x}_\alpha-{x}_{k})}{[(x_\alpha-x_k)^2+y_0^2+z_k^2]^{5/2}}\nonumber\\
&\ \ \ \times\frac{({x}_\beta-{x}_{k})}{[(x_\beta-x_k)^2+y_0^2+z_k^2]^{5/2}}\\
&\ \ \ \times \frac{t_k}{1+4\pi^2t_k^2f^2},\nonumber
\end{align}
with $\chi_0\equiv \frac{9e^2}{2\epsilon^2}\frac{y_0^2l^2}{(4\pi\epsilon_0)^2}$ and $\alpha,\beta=L,R$. In the limit where we average over all the possible distributions of charge fluctuators, the summation over ensembles and over fluctuators can be replaced by integrals with a corresponding probability distribution for their positions and switching times. We assume a uniform probability distribution for the position of the TLSs equal to the density $n_c$, and a distribution $\rho(t)$ for the switching time. Then, the average for the correlator of electric fields is:
\begin{align*}
\braket{C_{E_\alpha^xE_\beta^x}(f)}=\chi_0 n_c\int_{-\infty}^{\infty}dx&dz\frac{({x}_\alpha-{x})}{[(x_\alpha-x)^2+y_0^2+z^2]^{5/2}}\\
& \times\frac{({x}_\beta-{x})}{[(x_\beta-x)^2+y_0^2+z^2]^{5/2}}\\
& \times \int_0^\infty dt\ \rho(t) \frac{t}{1+4\pi^2t^2f^2}.
\end{align*}
For the auto-correlations, the integral in space can be solved analytically, obtaining:
\begin{align}
\braket{S_{E_\alpha^x}(f)}= \frac{\chi_0n_c\pi}{24y_0^6} G(f),
\label{eq:autocorr_analytic}
\end{align}
where we defined $G(f)\equiv\int_0^\infty dt\ \rho(t) t/({1+4\pi^2t^2f^2})$. For the cross-correlation, we could evaluate the leading order in an expansion of the integrand in powers of $y_0/d$ around $x=\pm d/2$. We obtained:
\begin{align}
\braket{|C_{E_L^xE_R^x}(f)|}= \frac{\chi_0n_c\pi}{192y_0d^5} G(f)+\mathcal{O}\left(d^{-7}\right).
\label{eq:crosscorr_analytic}
\end{align}
Equations~\eqref{eq:autocorr_analytic} and \eqref{eq:crosscorr_analytic} give:
\begin{align}
\bar{r}=\frac{\braket{|C_{LR}(f)|}}{\sqrt{\braket{S_L(f)}\braket{S_R(f)}}}=\frac{1}{8}\frac{y_0^5}{d^5},
\label{eq:rbar}
\end{align}
where the functions $G(f)$ cancel out making the result independent of the specific time distribution $\rho(t)$. This result is plotted as a purple line in Fig.~\ref{fig:scaling}.\\

\section{Non-uniqueness of the distribution of fluctuators}\label{app:distribution}

\begin{figure}[htb]
\centering
\includegraphics[width=\columnwidth]{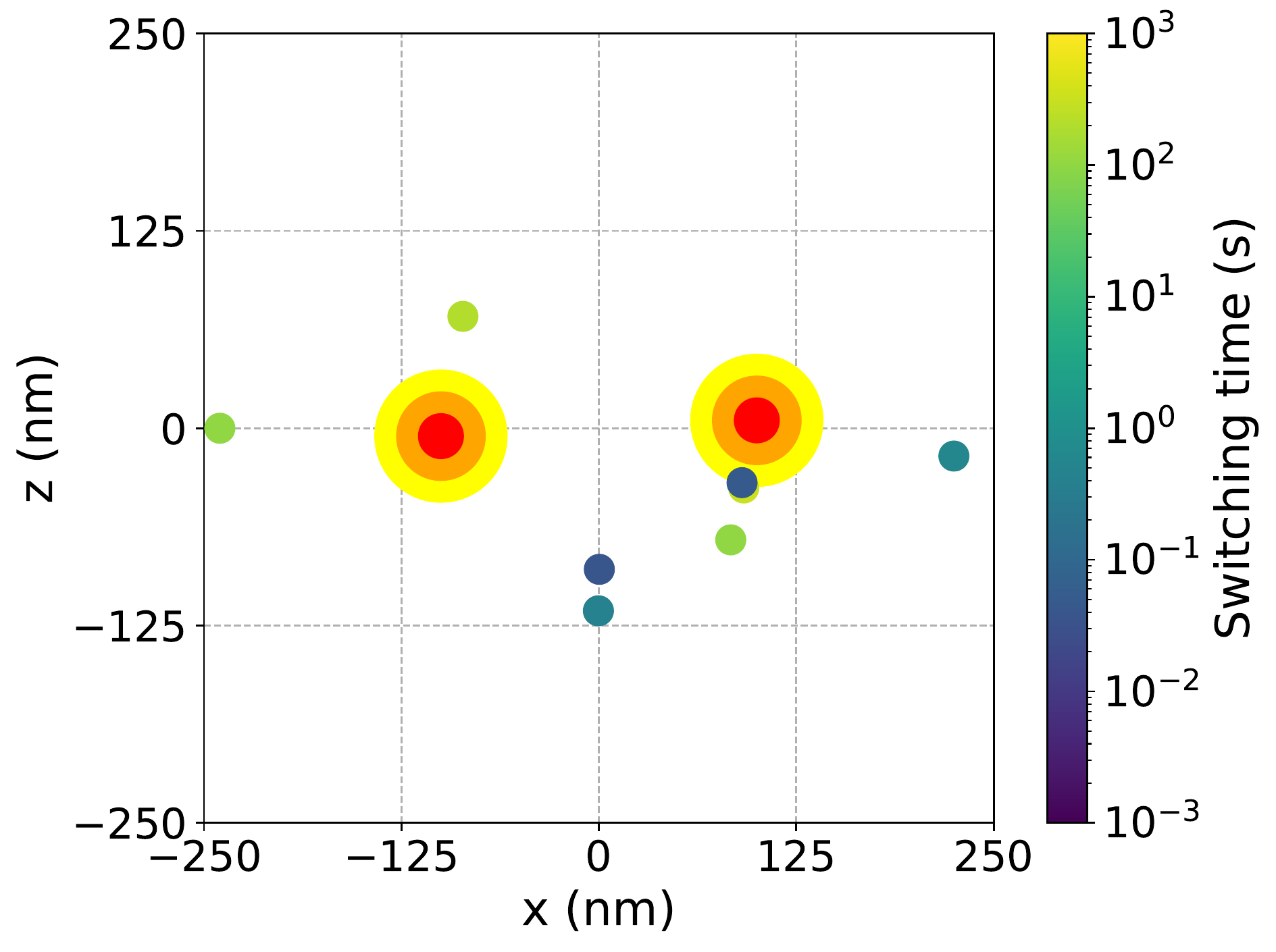}
\caption{Another ensemble of charge fluctuators that fits the measured data.}
\label{fig:charges_2}
\end{figure}

The distribution of TLSs that can reproduce our results is not unique. Figure~\ref{fig:charges_2} depicts a different ensemble of fluctuators that also leads to a satisfactory fit of the result from Fig.~\ref{fig:meas}.

\acknowledgments

We thank Á. Gutiérrez-Rubio for helpful discussion and comments on the manuscript. We thank the Microwave Research Group in Caltech for technical support. This work was supported financially by Core Research for Evolutional Science and Technology (CREST), Japan Science and Technology Agency (JST) (JPMJCR15N2 and JPMJCR1675), MEXT Quantum Leap Flagship Program (MEXT Q-LEAP) grant No. JPMXS0118069228, JST Moonshot R\&D Grant Numbers JPMJMS2065 and JPMJMS226B, JSPS KAKENHI grant Nos. 16H02204, 17K14078, 18H01819, 19K14640, 20H00237, and 21K14485, and Suematsu Fund, Advanced Technology Institute Research Grants. T.N. acknowledges support from JST PRESTO Grant Number JPMJPR2017, and J.Y. from JST PRESTO Grant Number JPMJPR21BA.

\bibliography{references}

\end{document}